\RequirePackage{fix-cm}
\documentclass[smallextended]{svjour3}       
\smartqed  
\usepackage{graphicx}
\usepackage{amsmath}
\begin{document}

\title{Emergence of space and cosmic evolution based on entropic force}


\author{Fei-Quan Tu \and Yi-Xin Chen}


\institute{Fei-Quan Tu \and Yi-Xin Chen, \\Zhejiang
Institute of Modern Physics, Zhejiang University,
Hangzhou, 310027, P. R. China\\
\email{fqtuzju@foxmail.com; yixinchenzimp@zju.edu.cn}
}

\date{Received: date / Accepted: date}

\maketitle

\begin{abstract}
In this paper, we propose a model in which an additional pressure due to the effects
of the entropic force is added to the ideal fluid. Furthermore, we obtain
the dynamic equation in the FRW universe which contains the quantum gravitational
effects based on the description of entropic force and emergence of
space. Our model can well explain the age of the universe and the effect of
the current accelerating expansion. We give the relation between the luminosity
distance and the redshift factor, and compare this relation with
that of lambda cold dark matter model($\Lambda CDM$
model).
\keywords{Entropic force, Emergence}
\end{abstract}

\section{Introduction}
\label{intro}
In a variational approach to gravity, we usually choose the Einstein-Hilbert(EH)
action as the fundamental action, which is an integral over the invariant
four-volume of a scalar curvature. In order to obtain the Einstein
equation, one usually add a Gibbons-Hawking surface term\cite{key-1} to
cancel the surface term obtained by the variation of the EH action. However, it
was pointed out that the true dynamical degrees of freedom of gravity
which make gravity intrinsically holographic reside in the surface
term rather than in the bulk term \cite{key-2}, and the EH action
possesses a holographic relation between the surface and bulk terms\cite{key-3,key-4}.
More significantly, the Einstein equation can be obtained just from
the surface term of the EH action\cite{key-2} and also from
the surface term of the more general actions in the case of pure gravity\cite{key-5}.
Therefore, the surface term is so important  that we can not disregard
it.

A large number of cosmological observations have showed that our universe
is in accelerating expansion. The fact is usually explained by the $\Lambda CDM$
model which is well in line with the data of astronomical observations
and implies that 73\% of the total energy in the universe is dark
energy. However, since Verlinde\cite{key-6}pointed out that gravity
could be explained as an entropic force caused by changes in the information
associated with the positions of material bodies, it was proposed
by Easson et al. that a driving term should be added to the Friedmann
dynamic equation because of the surface term obtained by the variation of the EH action
\cite{key-7}. In this entropic force scenario,
called ``entropic cosmology'', the expansion of the universe is
caused by the entropy of the surface. Based on the idea of entropic
cosmology, cosmic expansion were  investigated further in Refs.\cite{key-8,key-9,key-10}.

Padmanabhan\cite{key-11}suggested that cosmic space is emergent
as cosmic time progresses. He argued that the difference between the
number of the surface degrees of freedom and that of the bulk degrees
of freedom in a region of space drives the accelerating expansion
of the universe through a simple equation $\Delta V=\Delta t(N_{sur}-N_{bulk})$,
where $V$ is the Hubble volume in Planck units and $t$ is the cosmic
time in Planck units. Further, he derived the standard Friedmann equation
of the FRW universe. Emergent perspective of
gravity has been further investigated\cite{key-12,key-13,key-14,key-15}.

In this paper, we are interested in how to obtain the dynamic equation
without the cosmological constant. In other words, what may be the
origin of the ``dark energy''. The entropic force caused by the
surface term can lead to a negative pressure which causes the expansion
of the universe in the entropic force scenario. Hence we propose that
an additional pressure  is added to the matter which is taken as ideal
fluid. Furthermore, we obtain
the number of modified degrees of freedom in the bulk and also get
the corresponding dynamic equation in the FRW universe from the idea that
space is emergent as cosmic time progresses. This is a dynamic equation
which contains the quantum gravity effects. Then we obtain the modified
continuity equation. In order to  see the properties of the evolution
of the universe more clearly,
we obtain the solutions at the early universe and the present
epoch without considering the quantum effects.
The current rate of growth $a(t)=t^2$ is consistent with supernova observations.
We can explain the current accelerating expansion without the cosmological constant.
Quantum correction term of the dynamic equation in our model may cause inflation at the very early universe.
Therefore, the entropic force caused by the surface term may be explained as the
``dark energy''.  In the paper the units are chosen with
$c=\hbar=k_{B}=1$ and the current Hubble constant $H_{0}$ is taken as $70km\cdot s^{-1}\cdot Mpc^{-1}$
when we calculate the age of the universe.

\section{Dynamic equation of Friedmann universe based on entropic force
and emergence of space}
\label{intro}
The effective action is taken as a starting point in any fundamental
theories from the viewpoint of modern physics. So does Einstein
gravity theory. In general, the surface term obtained by the variation of the EH action is canceled by
adding to a extra surface term
in order to obtain the Einstein equation. But the surface
term is so important that we can't cancel it as we point out in introduction.

In this paper, we take the EH action as our fundamental starting point, which is
\begin{align}
S_{EH}&=\frac{1}{2\kappa}\int d^{4}x\sqrt{-g}L_{EH}+\int d^{4}x\sqrt{-g}L_{m}\notag\\
&=\frac{1}{2\kappa}\int d^{4}x\sqrt{-g}R+\int d^{4}x\sqrt{-g}L_{m}.
\end{align}
As Padmanabhan has pointed out\cite{key-3,key-4}, $\sqrt{-g}L_{EH}$ can be written as a sum of $L_{bulk}$ and  $L_{sur}$
where $L_{bulk}$ is quadratic in the first derivative of the metric and $L_{sur}$ is a total derivative which leads to a surface term in this action. If we ignore the term $L_{sur}$, then the usual Einstein equation is obtained.
Furthermore, there exists a key nontrivial relation, called holographic relation, $L_{sur}=-\frac{1}{2}\partial_{\lambda}\left(g_{\mu\nu}\frac{\partial L_{bulk}}{\partial(\partial_{\lambda}g_{\mu\nu})}\right)$.

Varying this action with respect to the metric,
we get
\begin{equation}
R_{\mu\nu}-\frac{1}{2}Rg_{\mu\nu}=\kappa T_{\mu\nu}+S_{\mu\nu},
\end{equation}
where the term $S_{\mu\nu}$ is caused by the surface term.
Generally, this term is described by a delta function because
of the locality of the theory. However, it should be a global effect
from the above holographic viewpoint. In our paper, the global
effect is described by the entropic force.

In the de Sitter space with radius $l$, there are the Hawking temperature $T=1/(2\pi l)$
and the entropy $S=A/4$ on the horizon,
where $A = 4\pi l^2$ is the cosmological horizon area\cite{key-16,key-17}.
From this aspect, the cosmological horizon is very similar to
the black hole horizon. Based on such similarity, the relation
between the horizon and thermodynamics in black hole physics
is often extended to the cosmological model,
and some progress about thermodynamics of the cosmological horizon has been made
(for example, see Refs.\cite{key-18,key-19,key-20}). Besides, our universe should be described
by the quantum language(e.g., the wave function of the universe
which obeys the Wheeler-DeWitt equation\cite{key-21,key-22})
from the point of view of the modern physics.
Based on the above reasons, we employ the quantum corrected entropy of
the black hole horizon as the entropy of the cosmological horizon\cite{key-23,key-24}
\begin{equation}
S=\frac{A}{4L_{p}^{2}}+\alpha\ln\frac{A}{4L_{p}^{2}},
\end{equation}
where $A=4\pi r^2$ is the area of the horizon in which $r$ is the radius of the horizon,
$\alpha$ is a parameter and $L_{p}$ is the Planck length.
The entropic force is given by
\cite{key-9,key-10}
\begin{equation}
F_{e}=-\frac{dE}{dr}=-T\frac{dS}{dr}.
\end{equation}
Combining Eq.(3) with Eq.(4), we obtain the corresponding entropic pressure
\begin{equation}
P_{e}=\frac{F_{e}}{A}=-\frac{T}{2L_{p}^{2}r}-\frac{\alpha T}{2\pi r^{3}}.
\end{equation}

In order to understand the emergence of space, a specific version
of holographic principle was suggested by Padmanabhan\cite{key-11}.
He suggested that our universe is being driven towards holographic equipartition(the
number $N_{sur}$ of degrees of freedom on the surface equals to the
number $N_{bulk}$ of degrees of freedom in the volume), and the law of evolution of
the universe is given by
\begin{equation}
\frac{dV}{dt}=L_{p}^{2}(N_{sur}-N_{bulk}).
\end{equation}

The effective number $N_{bulk}$ of degrees of freedom in the volume
is determined by the equipartition law of energy, so $N_{bulk}=|E|/[(1/2)T]$.
We take the ideal fluid as the matter of the universe, $\rho$ and
$p$ represent the energy density and the pressure of
the ideal fluid respectively. Further, we propose that the additional pressure $P_{e}$
is added to the ideal fluid. Under this proposal, the Komar energy $E$ contained
inside the volume $V=\frac{4}{3}\pi r^{3}$ is taken as $|\rho+3(p+P_{e})|V$.

Taking the Hubble radius $r=1/H$ as the radius of the horizon
and $T=H/2\pi$ as the temperature of the horizon, we obtain
\begin{equation}
P_{e}=-\frac{H^{2}}{4\pi L_{p}^{2}}-\frac{\alpha H^{4}}{4\pi^{2}}.
\end{equation}
Furthermore, we obtain
\begin{equation}
N_{bulk}=-\frac{(4\pi)^{2}}{3H^{4}}(\rho+3p)+\frac{4\pi}{H^{2}L_{p}^{2}}+4\alpha.
\end{equation}

The number of degrees of freedom on the spherical surface is given
by\cite{key-13} $N_{sur}=4S$, where $S$ is the entropy of the horizon.
Therefore, the number of degrees of freedom on the surface is
\begin{equation}
N_{sur}=\frac{4\pi}{H^{2}L_{p}^{2}}+4\alpha\ln\frac{\pi}{H^{2}L_{p}^{2}},
\end{equation}

If the universe is holographic or equivalently speaking it is pure de Sitter,
then $N_{sur}=N_{bulk}$ and we can obtain
\begin{equation}
\frac{3\alpha H^{4}}{4\pi^{2}}(1-\ln\frac{\pi}{H^{2}L_{p}^{2}})=\rho+3p.
\end{equation}
As what has been pointed out by Padmanabhan\cite{key-11}, there are some considerable evidences
that our universe is asymptotically de Sitter rather than pure de Sitter.
Hence, when the universe turns into holographic,
its volume should be infinitely large, then the energy density
and the pressure of matter both tend to zero.
Besides, the Hubble constant $H\rightarrow0$ when the universe tends to pure de Sitter.
Therefore, whether $\alpha=0$ or $\alpha\neq0$, Eq.(10) shows physical consistency.

Since our universe is asymptotically holographic equipartition,
using the law of evolution of the universe Eq.(6), we obtain
\begin{equation}
\frac{\ddot{a}}{a}=H^{2}+\dot{H}=-\frac{4\pi L_{p}^{2}}{3}(\rho+3p)+H^{2}+\frac{\alpha H^{4}L_{p}^{2}}{\pi}(1-\ln\frac{\pi}{H^{2}L_{p}^{2}}).
\end{equation}
This is the dynamic equation which contains the quantum gravitational effects.
It describes globally the expansion of the universe, and
it should be treated as the effective dynamic equation of the FRW universe.
In $\Lambda CDM$ model, the cosmological constant $\Lambda$
satisfies $\Lambda L_{p}^{2}\approx10^{-122}$\cite{key-25},
that is $\Lambda\approx10^{-52}m^{-2}$. At present, the Hubble constant
satisfies $H_{0}^{2}\approx10^{-53}m^{-2}$. Hence, the value of the cosmological
constant is consistent with that of the dominated term $H^{2}$ at
this epoch.  As we know, the Hubble constant $H$ is very great when time tends to zero.
Therefore, at the very early universe,
the quantum correction term $\frac{\alpha H^{4}L_{p}^{2}}{\pi}(1-\ln\frac{\pi}{H^{2}L_{p}^{2}})$
plays the dominating role on the cosmic dynamic evolution
and may cause the inflation of the universe.

\section{Properties of expansion of the universe}
\label{intro}
As usual, the continuity equation is satisfied, but its form is
$\dot{\rho}+3H(\rho+p+P_{e})=0$ because of the additional pressure
caused by the entropic force.
Here, we would like to give the physical interpretation. In the usual
gravity theory, the whole spacetime in which the horizons don't exist
is continue, so the continuity equation is $\dot{\rho}+3H(\rho+p)=0$.
However, there exists an Hubble horizon in our model.
Due to the holographic relation, the degrees
of freedom in the bulk can be described by that of the surface
equivalently. In turn, the existence of the horizon (the surface term)
effects the distribution of degrees of freedom in the bulk. The effects are
described by the additional pressure in our model.
Therefore, the effects can be described by
the modified continuity equation $\dot{\rho}+3H(\rho+p+P_{e})=0$
equivalently.
Substituting Eq.(7) into the continuity
equation, we have
\begin{equation}
\dot{\rho}+3H(\rho+p)=\frac{3H^{3}}{4\pi L_{p}^{2}}+\frac{3\alpha H^{5}}{4\pi^{2}}.
\end{equation}

We would like to make some remarks to explain the advantages of our model.
In the paper \cite{key-7}, the authors have pointed out that their model is best viewed as a phenomenological model
in which the surface term is introduced without rigorous derivation. Here we derive the results unambiguously
just by employing the idea of emergence and the assumption of
the entropic force. Moreover, our model contains the quantum effects.

In this section, in order to see some properties of expansion of the
universe more clearly, we consider $\alpha=0$, that is, we don't consider the
quantum effects. Then we obtain the dynamic equation of cosmic evolution
\begin{equation}
\frac{\ddot{a}}{a}=H^{2}+\dot{H}=-\frac{4\pi L_{p}^{2}}{3}(\rho+3p)+H^{2}
\end{equation}
and the continuity equation
\begin{equation}
\dot{\rho}+3H(\rho+p)=\frac{3H^{3}}{4\pi L_{p}^{2}}.
\end{equation}
Using the equation of state of the ideal fluid
\begin{equation}
p=\omega\rho,
\end{equation}
where $\omega$ is a non-negative parameter. By solving the above
three equations, we get the solutions $\dot{H}=-H^{2}$ and
$\dot{H}=-\frac{1+3\omega}{2}H^{2}$.

\subsection{Evolution of the universe under the solution $\dot{H}=-H^{2}$}

As we have seen, the solution $\dot{H}=-H^{2}$ doesn't have the parameter $\omega$, which implies that
the rate of expansion of the universe is the same at the different
epoches of the universe. We can further obtain the solution of the scale factor
\begin{equation}
a(t)=t.
\end{equation}
The energy density is $\rho=\frac{3}{4\pi(1+3\omega)L_{p}^{2}t^{2}}$,
which is related with the parameter $\omega$. The age of the universe in this case is
\begin{equation}
t_{0}=\frac{1}{H_{0}}=9.8\times10^{9}yr.
\end{equation}
The data of the supernova analyzed by the Supernova Cosmology Project
indicate that the age of universe is $t_{0}=13.4_{-1.0}^{+1.3}\times10^{9}yr$\cite{key-26}.
The age obtained in Eq.(17) is greater than that of the standard model
without the cosmological constant, but still less than the age obtained by the observation.

The general formula of the luminosity distance is given by the Ref.\cite{key-27}
\begin{equation}
d_{L}=\frac{1+z}{H_{0}}\int_{1}^{1+z}\frac{dy}{H/H_{0}},
\end{equation}
where $H_{0}$ is the current Hubble constant, and $z$ is the redshift factor defined by $z+1\equiv y=a_{0}/a$ in which $a_{0}$ is the current scale factor.
Then we can obtain $\frac{H}{H_{0}}=\frac{a_{0}}{a}=y$ by Eq.(16).
Substituting this result into Eq.(18), we have the relation
between the luminosity distance and the redshift factor:
\begin{equation}
H_{0}d_{L}=(1+z)\ln(1+z).
\end{equation}

For $\Lambda CDM$ model, the luminosity distance of the flat universe
is given by the Ref.\cite{key-28}. Its form is
\begin{equation}
H_{0}d_{L}=(1+z)\int_{0}^{z}dz^{'}[(1+z^{'})^{2}(1+\Omega_{m}z^{'})-z^{'}(2+z^{'})\Omega_{\Lambda}]^{-1/2},
\end{equation}
where $\Omega_{m}=\frac{\rho_{m}}{\rho_{c}}$ and $\Omega_{\Lambda}=\frac{\rho_{\Lambda}}{\rho_{c}}$. $\rho_{c}$
represents the critical density, $\rho_{m}$ and $\rho_{\Lambda}$ represent the density for matter and the
cosmological constant respectively.
It has been found that the description of the universe for $\Omega_{m}=0.27$ and $\Omega_{\Lambda}=0.73$
is consistent with  the data obtained by Wilkinson Microwave Anisotropy Probe(WMAP).

\begin{figure}
\centering
\includegraphics[width=8cm]{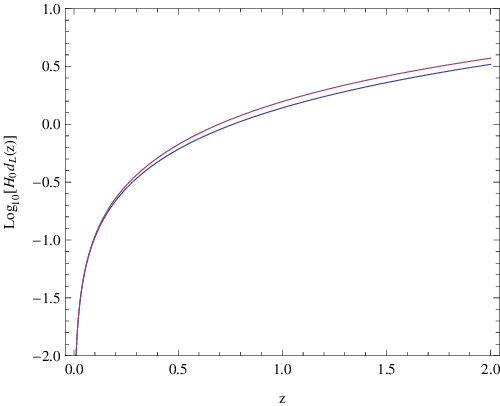}
\caption{(color online). The upper line represents the relationship
in $\Lambda CDM$ model, the lower line represents the relationship under the
solution $\dot{H}=-H^{2}$ in our model. Our result is well in line
with that of $\Lambda CDM$ model.}
\end{figure}

The deceleration parameter is defined as
\begin{equation}
q\equiv-\frac{\ddot{a}a}{\dot{a}^{2}}.
\end{equation}
Substituting Eq.(16) into Eq.(21), we obtain the deceleration parameter
$q=0$ in the solution $\dot{H}=-H^{2}$.

\subsection{Evolution of the universe under the solution $\dot{H}=-\frac{1+3\omega}{2}H^{2}$}

In this case, the solution is related with the parameter $\omega$,
which implies that the rate of expansion of the universe is different
at the different epoches of the universe. We will discuss
the results at the early time and the late time of the universe separately.

(i). At the early time of the universe, the radiation dominates the universe
at high energy scales, and the parameter $\omega=1/3$. Hence the evolution
law of the universe is determined by $\dot{H}=-H^{2}$. We can get
the solutions $H(t)=\frac{1}{t}$, $a(t)=t$ and the energy density
$\rho=\frac{3}{8\pi L_{p}^{2}t^{2}}$.

(ii). At the late time of the universe, the matter dominates the universe,
and the parameter $\omega=0$. Hence the evolution law of the universe
is determined by $\dot{H}=-\frac{1}{2}H^{2}.$ We can get the solution
$H(t)=\frac{2}{t}$, so we have
\begin{equation}
a(t)=t^{2},
\end{equation}
which varies with the square of time. Such a rate of growth is consistent with supernova observations \cite{key-29}. The energy density is $\rho=\frac{3}{2\pi L_{p}^{2}t^{2}}$.

Under the solution $\dot{H}=-\frac{1+3\omega}{2}H^{2}$, the age of universe $t_{0}$ satisfies
\begin{equation}
9.8\times10^{9}<t_{0}<19.6\times10^{9}yr,
\end{equation}
it is compatible with the astro-observational result. Using the Eq.(18), we
obtain the relation between luminosity distance and the redshift
factor
\begin{equation}
H_{0}d_{L}=z.
\end{equation}

\begin{figure}
\centering
\includegraphics[width=8cm]{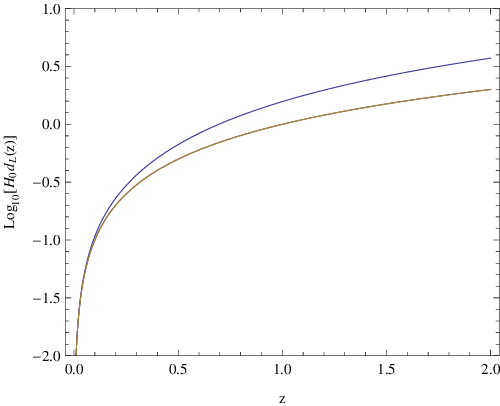}
\caption{(color online). The upper line represents the relationship
in $\Lambda CDM$ model, the lower line represents the relationship under
the solution $\dot{H}=-\frac{1}{2}H^{2}$ in our model. Our result has some deviation from that of $\Lambda CDM$ model.}
\end{figure}

Substituting Eq.(22) into Eq.(21), we obtain the deceleration
parameter $q=-\frac{1}{2}$. It shows that our universe is in accelerating
expansion even if the quantum correction isn't considered. The result
is consistent with that of the astro-observation.

\subsection{Remarks of the expansion of the universe}

In this section, we have analyzed the solutions of the expansion of the
universe without the quantum effects in our model.
We obtain the relation between the luminosity
distance and the redshift factor, the age of the universe and the deceleration
parameter. From these figures showing the relationship
between the luminosity distance and the redshift factor, we find that there is a better description of the astronomical observational data, but the effect of accelerating expansion doesn't exist under the solution $\dot{H}=-H^{2}$.
On the other hand, the relationship between the luminosity distance and
the redshift factor has some deviation from the results of
$\Lambda CDM$ model, but the effect of accelerating expansion is
obvious under the solution $\dot{H}=-\frac{1}{2}H^{2}$.

In order to obtain the correct age of the universe, we have to make the law of
evolution of the universe exist a transformation from $\dot{H}=-H^{2}$
to $\dot{H}=-\frac{1}{2}H^{2}$ at the epoch dominated by the matter. Thus we find
that the transformation of the law of  evolution occurs at $t=6.2\times10^{9}yr$
if the age of the universe $t_{0}$ is taken as $13.4\times10^{9}yr$.
That is, the universe has an evident accelerating expansion from then on. This is compatible with
the model of ``big bang'', which says the universe started accelerating expansion several billion years ago.
If it doesn't have a
transformation at that time, then it is hard to explain why we have two solutions at the late time
in our model.
Is this just an algebraic accident which doesn't have a physical explanation?
As for the reason for the transformation, it is beyond this paper, we will not discuss it.

\section{Conclusion}
\label{intro}
In this paper, our starting point is based on the EH action.
A surface term can be produced by varying this action with respect to the metric.
The surface term should cause a global effect which is described by the entropic
force from the holographic
viewpoint \cite{key-9}. In our model, we propose that an additional pressure
caused by the effects of the entropic force
is added to the matter which is taken as the ideal fluid
. Thus it leads to a change of the Komar energy of the
matter. Further, we obtain  the number of modified degrees of freedom
in the bulk. Based on the idea that space is emergent as cosmic time progresses
proposed by Padmanabhan, we obtain the corresponding dynamic equation
in the FRW universe. The dynamic equation describes globally the expansion
of the universe. Moreover, it contains the quantum correction term.

Then we analyze the properties of the expansion of the universe.
Using the dynamic equation of the evolution of the universe,
the modified continuity equation and the equation of state of the
ideal fluid, we obtain these solutions of the expansion of the universe
without the quantum correction. By analyzing the solutions of the epoch dominated by
the matter and the age of the universe, we find that it has a transformation
from the solution $\dot{H}=-H^{2}$
to the solution $\dot{H}=-\frac{1}{2}H^{2}$ at $t=6.2\times10^{9}yr.$
Thus we have a good description about the age of the universe and the effect of the current
accelerating expansion, providing an expansion for the dark energy in terms of entropic pressure.
The current rate of growth $a(t)=t^2$ is consistent with supernova observations.
The curve of the relation between the luminosity distance and the redshift
factor in our model has some deviation from the result of $\Lambda CDM$ model at high redshift region,
but it is understandable since the quantum correction term isn't considered in above discussions.
We also find that the quantum correction term plays a leading role  in the evolution of
the universe and may cause inflation at the very early time.
At the end of this paper, it should be stressed that the
solutions we obtain don't contain the quantum gravity effects.
The properties of cosmic evolution with the quantum gravity effects are worthy of further investigation. Our results are useful for
further understanding of the ``dark energy'' and the properties of the evolution of the universe.

\begin{acknowledgements}
This work is supported in part by the NSF of China Grant No. 11075138 and No. 11375150.
\end{acknowledgements}




\end{document}